# Characteristics of a Titanium Manganese redox flow battery based on Comsol

Anupam Saha
Dept. of Materials Science and Engineering
Khulna University of Engineering and Technology
Khulna, Bangladesh
anupam.kuet.mse@gmail.com

Shinthia Binte Eskender
Dept. of Materials Science and Engineering
Khulna University of Engineering and Technology
Khulna, Bangladesh
sinthiaoyshi@gmail.com

*Abstract*— **A simulation model and design of Titanium Manganese Redox Flow Battery (TMRFB) is proposed to study the distribution of dissociation rate, overpotential, current density, and electrode potential. TMRFB is one of the most promising new energy storages because of its high capacity and eco-friendly characteristics in the current condition of energy scarcity and environmental pollution. Moreover, Mn-based flow batteries are gaining popularity due to their inexpensive cost and high energy density in lieu of all vanadium redox flow batteries which are expensive. This research shows that the surface dissociation rate of $Ti^{4+}/Ti^{3+}$ and $Mn^{3+}/Mn^{2+}$ ions are higher at the membrane and lower at the inlet where the velocity of the electrolyte flow is higher; Furthermore, our work reveals that when the thickness of the electrode is compressed from 4.5 mm to 3 mm, overpotential reduces whereas current density and electrode potential increases. The COMSOL Multiphysics® software is used to solve the model's equations using the finite-element approach. From the dissociation rate it is concluded that less potential is required at the membrane for the oxidation-reduction reaction and with optimized electrolyte flow rate battery performance can be improved. Thus, electrode compression increases conductivity and battery performance.**

I. INTRODUCTION

In recent times, the use of flow batteries has revolutionized electric power companies, electric factories, offices & buildings because of their high-power density, low cost, and environment-friendly characteristics. A flow battery refers to an electrochemical device that directly generates electricity stored in materials by a chemical reaction. It works similarly to the conventional battery and fuel cells. One of the major applications of the flow battery system is the storage of electrical energy generated from distributed renewable sources. Wind and solar power are examples of renewable energy sources which has been quickly expanding all across the world. The widespread deployment of such power sources, whose outputs vary based on weather conditions, needs grid stability measures. Flow cell normally works on the dissolving of ions and their chemical reaction. Two liquid electrolyte dissolutions comprising dissolved metal ions as active masses are pumped to opposing ends of the electrochemical cell in redox flow batteries[1]. This battery has some characteristics like output power and capacity that can be designed in a variety of ways; higher energy efficiency; relatively higher reliability and stability; longer battery life.[2] But they are not ideal for tiny mobile equipment such as electric vehicles since their energy density is around one-tenth that of other varieties, including lithium-ion batteries.

A possible solution could be to use Titanium Manganese Flow Battery which is inexpensive and provides higher charge density. It is a Multiphysics coupling device with a sophisticated coupling system. The performance of the Titanium Manganese Flow Battery is heavily influenced by the electrochemical reaction, structure of the battery, transfer method of mass, and distribution of reaction area. Experimental research on the issues of strengthening and enhancement of performance of Titanium Manganese Flow Battery will need a significant amount of time and money[3]. A few diverse ways have been proposed and constructed to the improvement of the redox flow battery[4], [5]. Without taking into account the changes in the battery's parameters during the charge and discharge process, Li et al[6] constructed an AC impedance model that fails to accurately reproduce the battery's current characteristics and exterior voltage through charging and discharging processes. A 2D steady-state model was proposed by Yang et al [7] to show the simulation of the quality particle charge transfer coupling and the electrochemical reaction of Vanadium Redox Flow Battery (VRFB). The effects of the initial concentration of vanadium ions, initial concentration of $H^+$ ions, and current density are focused. The researcher analyzed different effects of electrolyte flow rate on electrodes and discussed the effects on the ohmic loss caused by the electrolyte and it can be improved. The impacts of different electrolyte flow rates on concentration distribution, influences on different properties by compressing electrodes, and effects of different electrode temperatures on electrode voltage are studied [8] based on the COMSOL software by following principles of VRFB. It is discovered that [9] using Ti with the Mn electrolyte to stabilize $Mn^{3+}$ ions and restrict particle development of $MnO_2$ might solve the main problem of solid $MnO_2$ precipitation owing to a disproportionation reaction.

However, there is a lack of analysis focusing on the depth analysis of the stack electrode process mechanism to analyze the electrode material impact and characteristics of electrochemical reaction on the performance of the battery. Herein, following the principles of Titanium Manganese Redox Flow Battery (TMRFB), the Nernst equation, and electrodynamics, a TMRFB simulation model is built using the COMSOL software. Then the model is used to optimize performance to improve programs, stack parameters, and examine under different conditions.

## II. THEORY OF TMRFB

The basic working principle of TMRFB is very close to that of the conventional battery charging-discharging scheme. Here, different oxidation states of Titanium ions work as the active material of the negative electrode and are stored in the negative electrolyte tank. As positive electrode material, Mn-ions at various oxidation states are used and stored in a positive electrolyte tank. To separate the positive and negative electrodes a membrane is used. The electrochemical reaction occurs during the flow of electrolytes over the electrode surface and the chemical energy is transformed into electric energy. The current is then collected and is driven through the bipolar plates[10].

During charging, reduction of $Ti^{4+}$ ($TiO^{2+}$) to $Ti^{3+}$ occurs at the negative electrode while oxidation of $Mn^{2+}$ to $Mn^{3+}$ occurs at the positive electrode through protons exchanging. Figure 1 shows the schematic diagram of TMRFB[11].

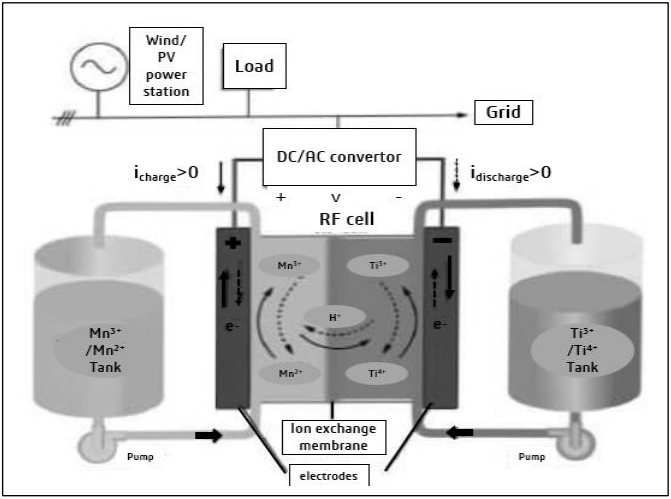

Fig. 1. The Schematic diagram of TMRFB

Reactions of positive and negative electrodes which respond electrochemically could be expressed as follows:

Positive electrode reaction:
$$Mn^{3+} + e^- \leftrightarrow Mn^{2+} \quad (1)$$
(E= 1.5 V)

Negative electrode reaction:
$$Ti^{3+} + H_2O \leftrightarrow TiO^{2+} + 2H^+ + e^- \quad (2)$$
(E= 0.1 V)

Cell reaction:
$$Ti^{3+} + Mn^{3+} + H_2O \leftrightarrow TiO^{2+} + Mn^{2+} + 2H^+ \quad (3)$$
(E= 1.4 V)

Disproportionation Reaction:
$$2Mn^{3+} + 2H_2O \leftrightarrow Mn^{2+} + MnO_2 + 4H^+ \quad (4)$$

## III. MECHANISM OF TMRFB

In this paper, a 2D steady-state model of the TMRFB is given, which is based on the battery's operating mechanism. Concentration field, dissociation field, electric field, and reaction involving electrochemistry are also taken into consideration. The Nernst-Einstein relation for mobility was used to model the electrolyte flow. The mass transport of species in solution was modeled using the Nernst–Planck equation. The reaction process of TMRFB is profoundly comprehended, based on the geometrical parameters, the impact of material properties, and stack performance depending on working conditions. The responding flow-concentrated species interface is used to model the flow and mass movement. An electrolyte-electrode interface coupling is used to model the reaction.

### A. Assumptions of Modes

The primary assumption of modes involves:
- The three components of battery- electrolyte, electrode, and membrane are assumed to be isotropic as the 2D steady-state model of TMRFB is simulated,
- The membrane of the cell represents as a conductor of proton without taking additional ions into account,
- Unidirectional, incompressible, and laminar Electrolyte flow is assumed.
- Side reactions like water evolution, as well as consequent bubble production, are not taken into account.
- Species transport is defined using the dilute-solution approximation.

### B. Modeling of Electrodes

Nernst equations are used to calculate the electrode equilibrium potential:

$$E_{neg} = E_{0,neg} + \frac{RT}{F} ln\left\{\frac{\alpha_{Ti^{3+}}}{\alpha_{Ti^{4+}}}\right\} \quad (5)$$

$$E_{pos} = E_{0,pos} + \frac{RT}{F} ln\left\{\frac{\alpha_{Mn^{2+}} \cdot \alpha_{H^+}^2}{\alpha_{Mn^{2+}}}\right\} \quad (6)$$

The negative electrode reaction's reference potential is $E_{0,neg}$ and the positive electrode reaction's reference potential is $E_{0,pos}$, R defines molar gas constant, the temperature is defined by T, and F is referred to as Faraday's constant.

Butler-Volmer type kinetic expressions describe reaction processes of positive and negative electrodes:

$$i_{neg} = A.\,i_{0,neg}\left[e^{\frac{(1-\alpha_{neg})F\eta_{neg}}{RT}} - e^{\frac{(-\alpha_{neg})F\eta_{neg}}{RT}}\right] \quad (7)$$

$$i_{0,neg} = Fk_{neg}(\alpha_{Ti^{3+}})^{1-\alpha_{neg}}.(\alpha_{Ti^{4+}})^{\alpha_{neg}} \quad (8)$$

$$i_{pos} = A.\,i_{0,pos}\left[e^{\frac{(1-\alpha_{pos})F\eta_{pos}}{RT}} - e^{\frac{(-\alpha_{pos})F\eta_{pos}}{RT}}\right] \quad (9)$$

$$i_{0,pos} = Fk_{pos}(\alpha_{Mn^{3+}})^{1-\alpha_{pos}}.(\alpha_{Mn^{2+}})^{\alpha_{pos}} \quad (10)$$

Where A refers to the specific surface area, $\alpha_{neg}$ and $\alpha_{pos}$ denote the negative and positive transfer coefficients, respectively, and $k_{neg}$ and $k_{pos}$ denote the negative and positive reaction rate constants, respectively.

The overpotentials for the positive electrode and the negative electrode are as follows:

$$\eta_{neg} = \phi_{s,neg} - \phi_{l,neg} - E_{neg} \quad (11)$$

$$\eta_{pos} = \phi_{s,pos} - \phi_{l,pos} - E_{pos} \quad (12)$$

Where the $\varphi_s$= the electrode potential, and
$\varphi_l$ = the electrolyte potential.

Ohm's law gives the porous electrode current equation:
$$i_s = -\sigma_s^{eff}\nabla\phi_s \quad (13)$$

The Brueggemann correction is used to compute the effective conductivity $\sigma_s^{eff}$. From this equation, it can be seen porous media conductivity is related to the solid materials porosity ε,

$$\sigma_s^{eff} = (1-\varepsilon)^{3/2}\sigma_s \qquad (14)$$

When the thickness of the electrode is compressed, changes occur in the porosity ε of the electrode.

$$\varepsilon = 1 - \frac{d_0}{d}(1-\varepsilon_0) \qquad (15)$$

Where, the $\varepsilon_0$ = porosity of the electrode,
$d_0$ = electrode thickness before compression and
$d$ = electrode thickness after compression.
The electrode surface area will be modified keeping constant the actual surface area, hence the electrode's specific surface area, $\alpha$ will be affected.

$$\alpha = \alpha_0 \frac{d_0}{d} \qquad (16)$$

### C. Dissociation of $H_2SO_4$

The following processes are followed to dissociate sulfuric acid:

$$H_2SO_4 \leftrightarrow H^+ + HSO_4^- \qquad (17)$$
$$HSO_4^- \leftrightarrow H^+ + SO_4^{2-} \qquad (18)$$

$r_d$ describes the degree of dissociation:

$$r_d = k_d \left( \frac{\alpha_{H^+} \cdot \alpha_{HSO_4^-}}{\alpha_{H^+} + \alpha_{HSO_4^-}} - \beta \right) \qquad (19)$$

Where $k_d$ defines $HSO_4^-$ dissociation rate constant and β refers to the dissociation level. The $H^+$ concentration in the electrolyte is affected by the dissociation of sulphuric acid, hence $r_d$ is an important parameter.

### D. Current density determination

The molar flux $N_i$ of ions is calculated by the Nernst-Planck equation.

$$N_i = -D_i \cdot \nabla_{c_i} - Z_i \cdot u_{mob,i} \cdot F \cdot c_i \cdot \nabla_{\phi_l} + c_i \cdot u \qquad (20)$$

In this equation, diffusion flux is known from the 1st term, where $D_i$ denotes diffusion coefficient and the 2nd term refers to migration term. Migration term is calculated by the electrostatic charge of the ion $Z_i$ and mobility $u_{mob,i}$. 3rd term denotes as convection term where u refers to fluid velocity.
The current density of the electrolyte in a porous electrode is determined by Faraday's law. Here current density is the multiplication of total molar flux and the number of charges of the material.

$$i_l = F \cdot \sum_{i=1}^{n} Z_i \left( -D_i \cdot \nabla_{c_i} - Z_i \cdot u_{mob,i} \cdot F \cdot c_i \cdot \nabla_{\phi_l} \right) \qquad (21)$$

### E. Selection of Boundary

2nd current distribution interface boundary condition and tertiary current distribution interface boundary condition are built up in this model.
1st requirement for the selection of boundary is that the electrolyte current density should be the same as the membrane current density.

$$\vec{n} \cdot \overrightarrow{i_{l,e}} = \vec{n} \cdot \overrightarrow{i_{l,m}} \qquad (22)$$

Here electrolyte current density is denoted by $i_{l,e}$, and membrane current density is denoted by $i_{l,m}$.
The membrane current is related to proton flux since only protons H + cross the membrane.

$$n \cdot N_{+,e} = n \cdot \frac{i_{l,m}}{F} \qquad (23)$$

In the 2nd current distribution interface, the electrolyte potential limit is set up:

$$\phi_{l,m} = \phi_{l,e} + \frac{RT}{F} \ln\left\{\frac{\alpha_{+,m}}{\alpha_{+,e}}\right\} \qquad (24)$$

Here chemical activity of proton is denoted by $\alpha_{+,m}$

### F. Computational setting

COMSOL, a finite element analysis software, was used to build this battery model. And the model equations were solved using this software. It was necessary to set up the corresponding parameters, global definitions, model inputs, components, research physics, and boundary conditions. In COMSOL, a mesh with 4616 components was constructed by following the conventional size. The mesh density was higher towards the membrane where the main reaction takes place. Table 1 shows the parameter settings.

TABLE I. PARAMETERS FOR MODEL SETUP

| Parameter | Value |
|---|---|
| The thickness of electrode (d) | 0.0045 m |
| Temperature (T) | 293.15 K |
| Negative diffusion Coefficient ($D_{neg}$) | 4×10$^{-9}$ m$^2$/s |
| Positive diffusion Coefficient ($D_{pos}$) | 1×10$^{-10}$ m$^2$/s |
| Specific area (*a*) | 3.5×10$^5$ |
| Positive electrode standard potential ($E_{0, pos}$) | 1.5 V |
| Negative electrode standard potential ($E_{0, neg}$) | 0.1 V |
| Positive electrode transfer coefficient ($\alpha_{pos}$) | 0.5 |
| Negative electrode transfer coefficient ($\alpha_{neg}$) | 0.5 |
| Positive electrode rate constant ($k_{pos}$) | 5×10$^{-6}$ m/s |
| Negative electrode rate constant ($k_{neg}$) | 6.33×10$^{-8}$ m/s |
| Initial concentration of Ti$^{3+}$ ($C_{Ti}^{3+}$) | 220 mol m$^3$ |
| Initial concentration of Ti$^{4+}$ ($C_{Ti}^{4+}$) | 850 mol m$^3$ |
| Initial concentration of Mn$^{2+}$ ($C_{Mn}^{2+}$) | 850 mol m$^3$ |
| The initial concentration of Mn$^{3+}$ ($C_{Mn}^{3+}$) | 220 mol m$^3$ |

## IV. RESULT AND DISCUSSION

In figure 2, the blue gradient refers to a dissociation rate ranging from -10 to -20 and the red gradient refers to a dissociation rate greater than -5. It is seen that the surface dissociation rate of *Ti$^{4+}$/ Ti$^{3+}$ and Mn$^{3+}$/Mn$^{2+}$* ions are higher at the membrane and lower at the inlet where the velocity of the electrolyte flow is higher. It proves that when velocity is higher the reaction rate of the ions is lower and more potential is needed to activate the reactions. But far from the middle of the flow where velocity is lower, the reaction rate of the ions is higher and less potential is needed to activate the reactions. At the membrane, a more electrochemical reaction takes place with the optimized flow rate of the electrolyte. The model accurately reflects the internal state of the cell at the time of operation and this demonstrates that the flow field's design is crucial to the battery's optimization.

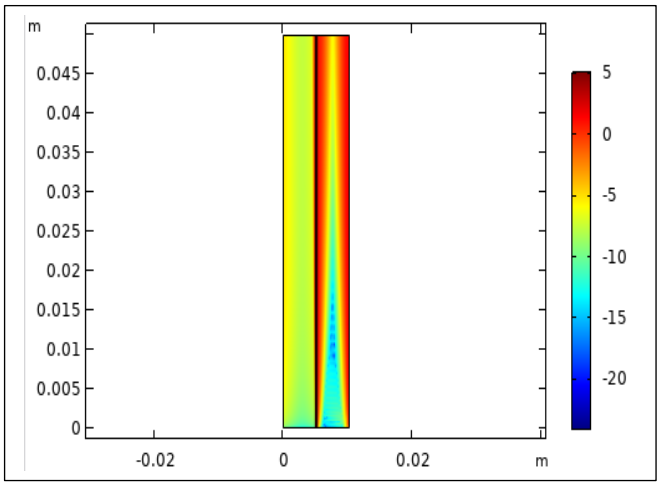

Fig. 2. Dissociation rate of $Ti^{4+}/Ti^{3+}$ and $Mn^{3+}/Mn^{2+}$ ions

From figure 3, figure 4, and figure 5, it is seen that the dissociation rate is lower at the maximum flow rate and it gradually increases. The flow rate of the electrolyte plays a vital role in the redox flow cell.

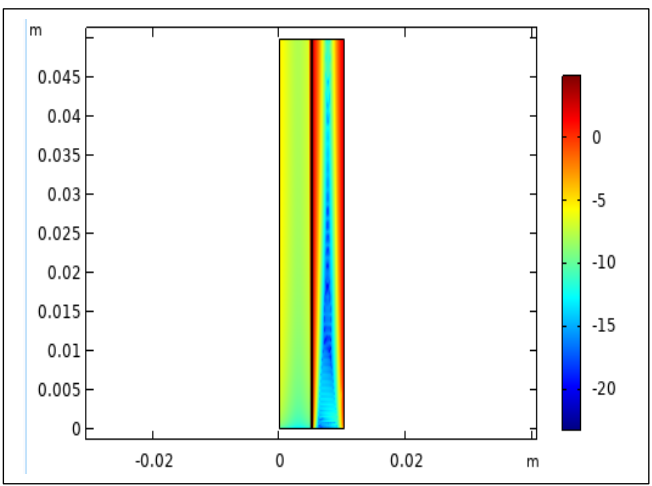

Fig. 3. Dissociation rate at flow rate 40 ml/min

In figure 3 the electrolyte flow rate is 40 ml/min, the dissociation rate of $Mn^{3+}/Mn^{2+}$ ions at the middle of flow are -11.513 $s^{-1}$ and it increases fast towards the membrane. At the membrane, the rate is 4.1277 $s^{-1}$.

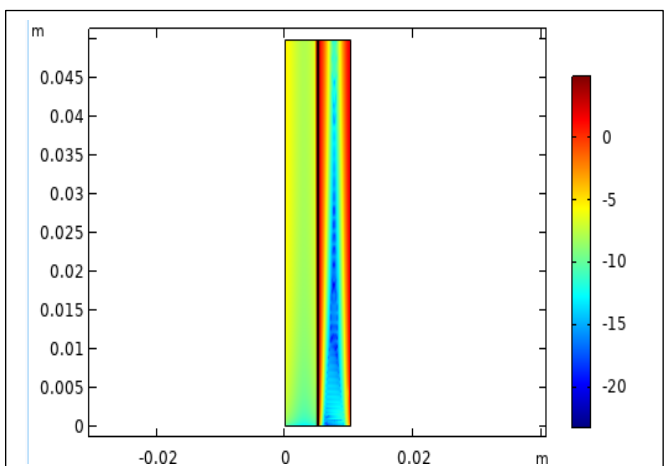

Fig. 4. Dissociation rate at flow rate 90 ml/min

In figure 4, at a flow rate of 90 ml/min the dissociation rate increases slowly. At the middle of inlet velocity, the value is -15.4526 $s^{-1}$ and slowly increases to 4.0608 $s^{-1}$ at the membrane.

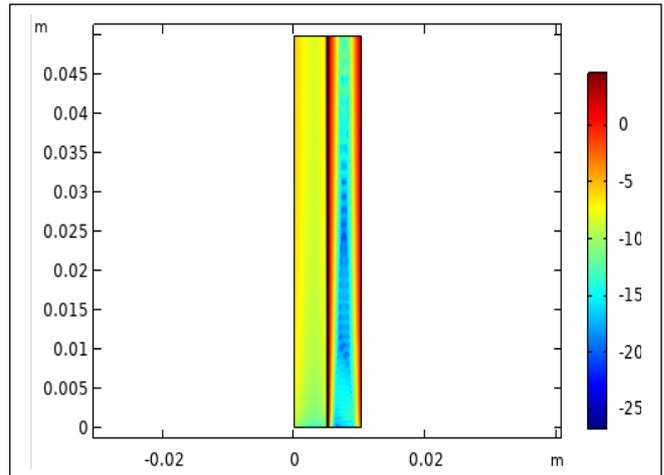

Fig. 5. Dissociation rate at a flow rate of 180 ml/min

In figure 5, the blue color gradient refers to a dissociation rate ranging from -15 to -25. For more flow rate of electrolyte, the ions don't get enough time to react and it dissociates less. Here, at an electrolyte flow rate, 180 ml/min dissociation rate is lower for a greater area. And it increases very slowly to 3.248 $s^{-1}$ at the membrane. From this dissociation rate, it is found that less potential is needed at the membrane for the oxidation-reduction reaction and the dissociation rate decreases with increasing electrolyte flow rate. For Ti-Mn cells, the increased rate of dissociation from inlet to membrane is higher.

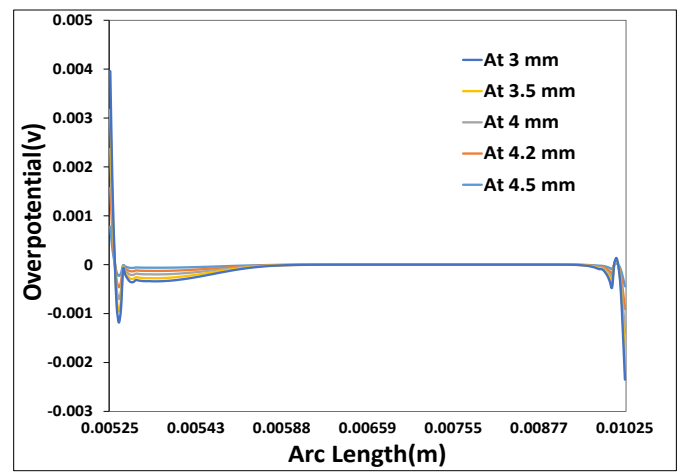

Fig. 6. Overpotential distribution with compression of electrode

The electrode was compressed while the electrolyte flow rate $v$ was held constant. After compression, the electrode's thickness is reduced from 4.5 mm to 3 mm. The electrode overpotential is calculated using the mechanism model's numerical analysis, as shown in Figure 6. In figure 6 it is seen that overpotential is varying at the membrane (0.00525 m) and outlet section (0.01025 m) but it is constant from arc length 0.0058m to 0.00877m. From this, it can be demonstrated that oxidation-reduction reaction occurs mainly at the membrane and the outlet. It is discovered that the enhancement in cell performance is mostly applicable to the decrease in ohmic loss induced by electrode compression, which leads to a reduction in porosity of electrode and as a result specific surface area and conductivity increase.

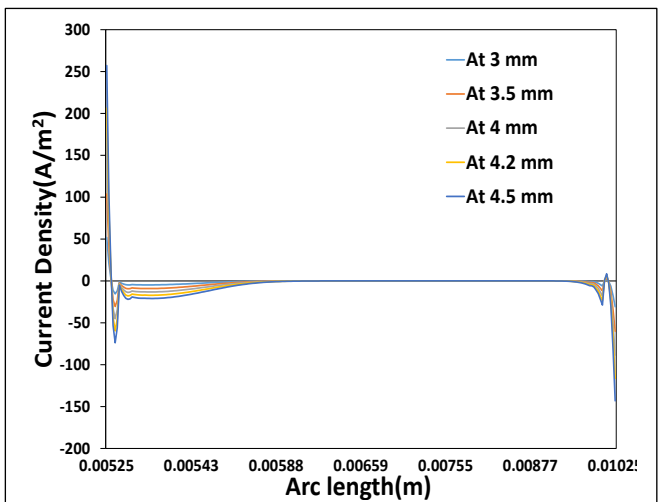

Fig. 7. Current density distribution with compression of electrode

Figure 7 shows that under compression of the electrode the current density gradually increases. At 4.5 mm the current density is the lowest and at 3 mm current density is maximum. From this, it can be seen that when the electrode is compressed, the overpotential decreases, and the flow of electrons increases. And it helps to increase the current density. As a result, conductivity increases with the compression of the electrode.

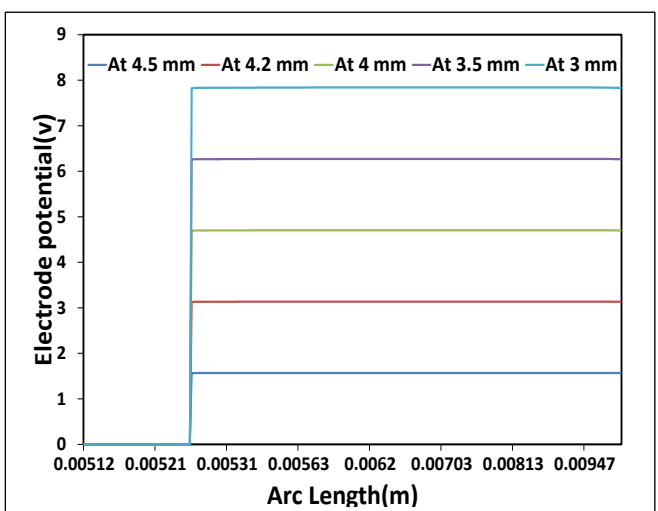

Fig. 8. Electrode potential distribution with compression of electrode

In figure 8, the electrode potential is constant up to 0.00521m. At the membrane (0.00525m), it increases instantly because of the oxidation-reduction reaction. It is found that under compression of the electrode, electrode potential increases. For the 4.5 mm electrode, the potential is lower and for 3 mm, it increases to a higher point. Electrode potential denotes the electron reduction and oxidation rate. With higher potential, the electron transfer rate is higher and it helps to increase the current flow. As a result, electron transfer rate and current flow increase with compression of the electrode. So, the performance of the battery is increased by the compression of the electrode.

## V. CONCLUSION

In this research, TMRFB is modeled and simulated. This simulation model is made to analyze the impacts of dissociation rate, the flow rate of electrolyte, and compression of the electrode on performance characteristics:

- From the dissociation rate of the Titanium Manganese ions, it can be concluded more potential is required to activate the reactions where velocity is more and less potential is required where velocity is lower. With an optimized electrolyte flow rate, more reaction occurs at the membrane.
- The extreme rate of flow is so rapid for the oxidation and reduction reaction, lowering the battery's efficiency. As a result, when the electrolyte flow rate is optimized, it is vital to consider the battery's performance loss.
- Compression of the electrode decreases the porosity, increases surface area, conductivity, and electron transfer rate. The terminal voltage is also reduced as ohmic loss is reduced and it increases current density. As porosity is reduced, electrode potential over the surface is increased. And with the increase of electrode potential, the oxidation-reduction rate increases. This indicates that the battery's performance has been improved.

REFERENCES

[1] Y. Yao, J. Lei, Y. Shi, F. Ai, and Y.-C. Lu, "Assessment methods and performance metrics for redox flow batteries," Nat. Energy, vol. 6, no. 6, pp. 582–588, 2021.

[2] H. Zhang, W. Lu, and X. Li, "Progress and Perspectives of Flow Battery Technologies," Electrochem. Energy Rev., vol. 2, no. 3, pp. 492–506, 2019, doi: 10.1007/s41918-019-00047-1.

[3] L. Qiao, C. Xie, M. Nan, H. Zhang, X. Ma, and X. Li, "Highly stable titanium-manganese single flow batteries for stationary energy storage," J. Mater. Chem. A, vol. 9, no. 21, pp. 12606–12611, 2021, doi: 10.1039/d1ta01147b.

[4] Y. Lv et al., "Recent advances in metals and metal oxides as catalysts for vanadium redox flow battery: properties, structures, and perspectives," J. Mater. Sci. Technol., vol. 75, pp. 96–109, 2021.

[5] A. Abbas et al., "Effect of electrode porosity on the charge transfer in vanadium redox flow battery," J. Power Sources, vol. 488, p. 229411, 2021.

[6] G. Li, Z.-W. Tang, H. Nie, and J. Tan, "Modelling and controlling of vanadium redox flow battery to smooth wind power fluctuations [J]," Power Syst. Prot. Control, vol. 22, pp. 115–119, 2010.

[7] W. W. Yang, Y. L. He, and Y. S. Li, "Performance Modeling of a Vanadium Redox Flow Battery during Discharging," Electrochim. Acta, vol. 155, pp. 279–287, 2015, doi: 10.1016/j.electacta.2014.12.138.

[8] M. Zhu, Q. Wu, X. Chi, and Y. Luo, "Simulation of all-vanadium redox flow batteries based on COMSOL," Proc. 29th Chinese Control Decis. Conf. CCDC 2017, pp. 6977–6982, 2017, doi: 10.1109/CCDC.2017.7978439.

[9] Y. Dong, H. Kaku, H. Miyawaki, R. Tatsumi, K. Moriuchi, and T. Shigematsu, "Titanium-manganese electrolyte for redox flow battery," SEI Tech. Rev., no. 84, pp. 35–40, 2017.

[10] R. Guidelli and G. Piccardi, "The voltammetric behaviour of the Mn2+, Mn3+, Mn4+ system in 15 N H2SO4 on a smooth platinum microelectrode," Electrochim. Acta, vol. 13, no. 1, pp. 99–107, 1968, doi: 10.1016/0013-4686(68)80010-6.

[11] P. Alotto, M. Guarnieri, and F. Moro, "Redox flow batteries for the storage of renewable energy: A review," Renew. Sustain. energy Rev., vol. 29, pp. 325–335, 2014.